\documentstyle[12pt,amsmath,amssymb,amsbsy,graphicx]{article}

\def\figwidth{0.5\textwidth}
\def\est{\varepsilon_F^*}
\def\vk{{\bf k}}
\def\v2o3{V$_2$O$_3$}
\def\uc2{$U_{c2}$}
\def\uc1{$U_{c1}$}

\begin{document}

\pagestyle{empty}

\fontfamily{ptm}

\selectfont

\vspace{1cm}
\begin{center}
\noindent {\Large \bf
A brief review of recent advances on the Mott transition:\\
unconventional transport,
spectral weight transfers, and critical behaviour}
\end{center}
\vspace{1cm}

\noindent {\bf A.~Georges$^a$,  S.~Florens$^b$ and T.A.~Costi$^b$}

\vspace{1cm}
\noindent
$^a${\it LPT - Ecole Normale Sup\'erieure 24, rue Lhomond 75231 Paris Cedex 05\\
and: CPHT - Ecole Polytechnique, 91128 Palaiseau Cedex;
e-mail: antoine.georges@cpht.polytechnique.fr}  \\
$^b${\it Institut f\"{u}r Theorie der Kondensierten Materie,
Universit\"{a}t Karlsruhe,
Postfach 6980,
76128 Karlsruhe;
e-mail: florens@tkm.physik.uni-karlsruhe.de (S.Florens),
tac@tkm.physik.uni-karlsruhe.de (T. Costi)}
\vspace{1cm}

\noindent \hfill  \parbox{16.5cm}
{\footnotesize {\bf Abstract.}
Strongly correlated metals close to the Mott transition display unusual
transport regimes, together with large spectral weight transfers in optics and
photoemission.
We briefly review the theoretical understanding of these effects, based
on the dynamical mean-field theory, and emphasize the key role played by the two
energy scales associated with quasiparticle coherence scale and with the Mott gap.
Recent experimental results on two-dimensional
organic compounds and transition metal oxides are considered in this perspective.
The liquid-gas critical behaviour at the Mott critical endpoint is also discussed.
Transport calculations using the numerical renormalization group are presented.
}

\vspace{.5cm}

\vspace{.2cm}
\noindent
{\it To appear in}: Proceedings of the Vth International Conference on Crystalline
Organic Metals, Superconductors and Magnets (ISCOM 2003) to be published in
Journal de Physique IV (EDP Sciences).

\vspace{.9cm}
\noindent
{\bf 1. MATERIALS ON THE VERGE OF THE MOTT TRANSITION}
\vspace{.4cm}

\begin{figure}
\begin{center}
\begin{tabular}{cc}
\includegraphics[width=8 cm]{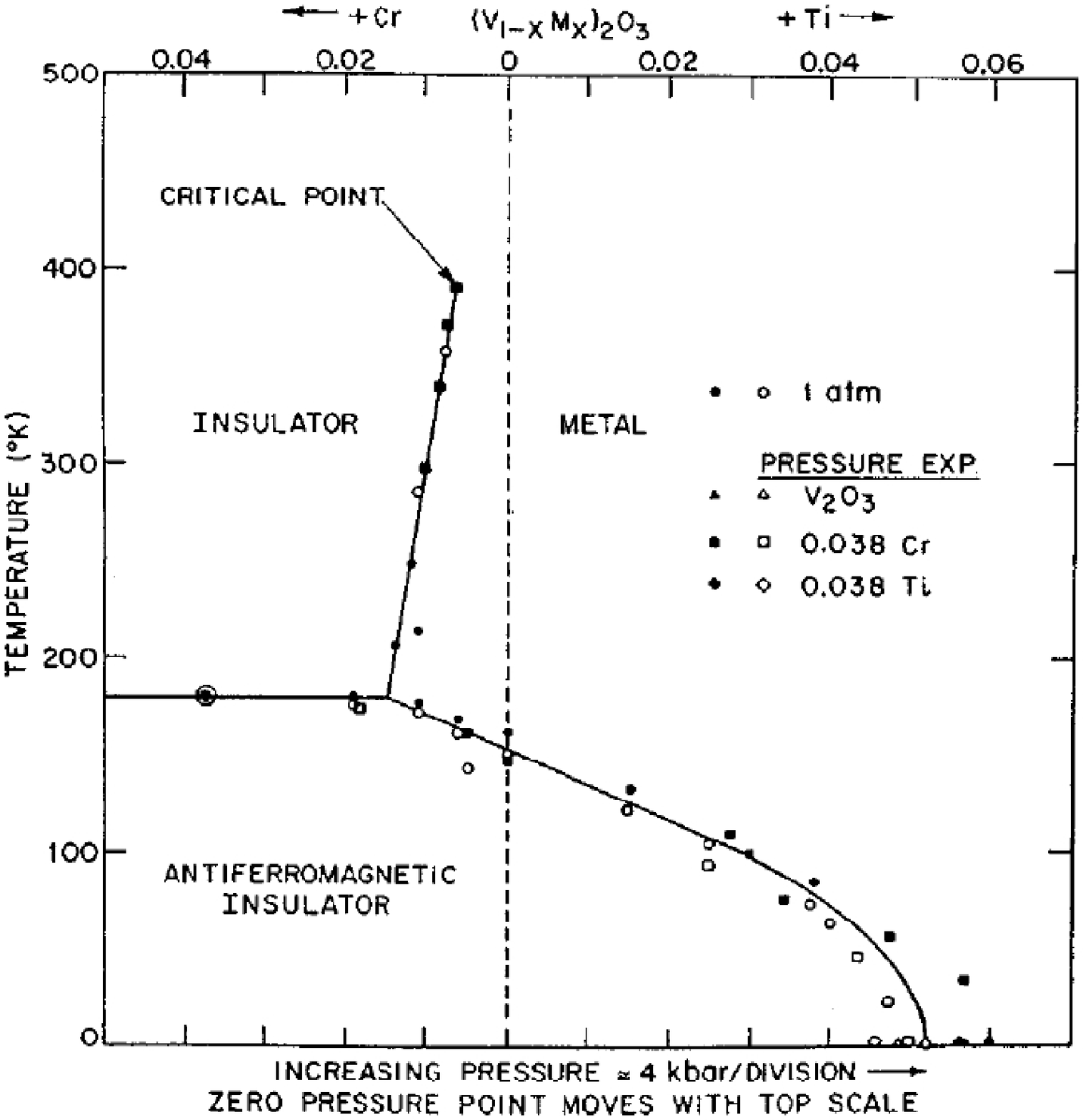} &
\includegraphics[width=9 cm]{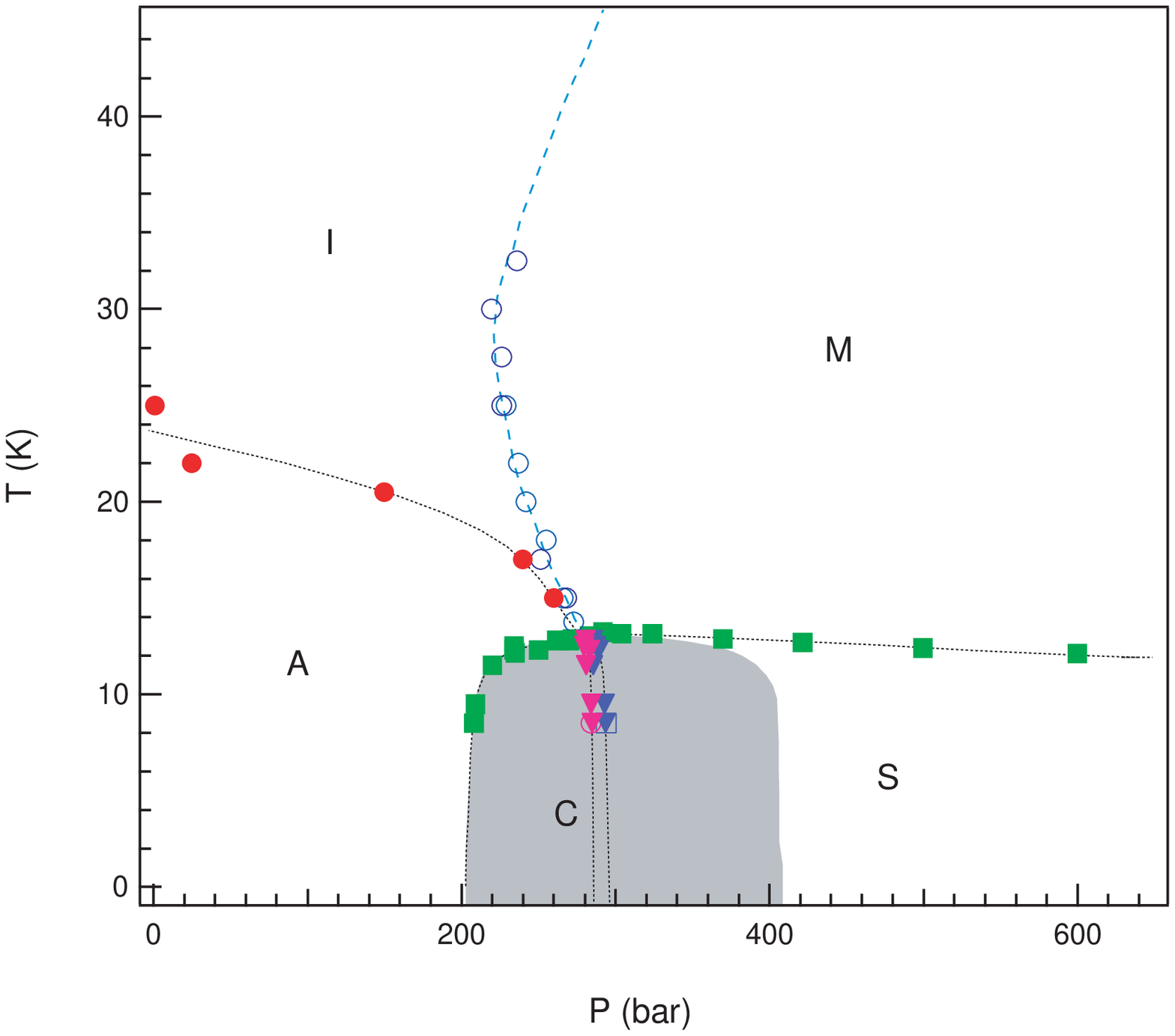}
\end{tabular}
\end{center}
\caption{
{\footnotesize
Phase diagram of (V$_{1-x}$ Cr$_{x}$)$_2$O$_3$ (left, after\cite{McWhan73}) and
of $\kappa$-(BEDT-TTF)$_{2}$Cu[N(CN)$_{2}$]Cl (right, after \cite{Lef00}),
as a function of Cr-concentration and pressure (left), and of pressure (right).
}}
\label{fig:materials}
\end{figure}

The Mott phenomenon - that interactions between electrons can be responsible for the
insulating character of a material -
plays a key role in the physics of strongly correlated electron materials.
Outstanding examples \cite{imada_mit_review} are transition-metal oxides (e.g superconducting cuprates),
fullerene compounds, as well as organic conductors.
A limited number of these materials are poised right on the verge of this electronic
instability. This is the case, for example, of V$_2$O$_3$, NiS$_{2-x}$Se$_x$ and of quasi two-dimensional
organic conductors of the $\kappa$-BEDT family. These materials are particularly
interesting for the fundamental investigation of the Mott transition, since they offer the
possibility of going from one phase to the other by varying some external parameter
(e.g chemical composition,temperature, pressure,...).
Varying external pressure is definitely
a tool of choice since it allows to sweep continuously from the insulating phase
to the metallic phase (and back). The phase diagrams of (V$_{1-x}$ Cr$_{x}$)$_2$O$_3$ and
of $\kappa$-(BEDT-TTF)$_{2}$Cu[N(CN)$_{2}$]Cl under pressure are displayed
in Fig.~\ref{fig:materials}.

There is a great similarity between the high-temperature part of the phase diagrams of these materials,
despite very different energy scales. At low-pressure they are {\it paramagnetic} Mott insulators, which
are turned into metals as pressure is increased. Above a critical temperature $T_c$
(of order $\sim 450$K for the oxide compound and $\sim 40$K for the organic one), this
corresponds to a smooth crossover. In contrast, for $T<T_c$ a first-order transition is observed, with a
discontinuity of all physical observables (e.g resistivity). The first order transition line
ends in a second order critical endpoint at $(T_c,P_c)$. We observe that in both cases, the
critical temperature is a very small fraction of the bare electronic
energy scales (for \v2o3 the half-bandwidth is of order $0.5-1$~eV, while it is of
order $2000$~K for the organics).

There are also some common features between the low-temperature part of the phase diagram
of these compounds, such as the fact that the paramagnetic Mott insulator orders into an
antiferromagnet as temperature is lowered. However, there are also striking differences: the
metallic phase has a superconducting instability for the organics, while this is not the case for
V$_2$O$_3$. Also, the magnetic transition is only superficially similar\,: in the case of
\v2o3, it is widely believed to be accompanied (or even triggered) by orbital
ordering\cite{bao_v2o3} (in contrast to NiS$_{2-x}$Se$_x$\cite{kotliar_V2O3_NiS}), and
as a result the transition is first-order. In general, there is a higher degree of universality
associated with the vicinity of the Mott critical endpoint than in the low-temperature region,
in which long-range order takes place in a material- specific manner.

Mott localization into a paramagnetic insulator implies a high spin entropy, which must therefore
be quenched in some way as temperature is lowered. An obvious possibility is magnetic
ordering, as in these two materials. In fact, a Mott transition between a paramagnetic Mott insulator
and a metallic phase is only observed in those compounds where magnetism is sufficiently {\it frustrated}
so that the transition is not preempted by magnetic ordering. This is indeed the case in both
compounds discussed here: \v2o3 has competing ferromagnetic and antiferromagnetic exchange constants,
while the two-dimensional layers in the organics have a triangular structure.
Another possibility is that the entropy is quenched through a Peierls instability (dimerization),
in which case the Mott insulator can remain paramagnetic (this is the case, for example, of VO$_2$).
Whether it is possible to stabilize a paramagnetic Mott insulator down to $T=0$ without breaking
spin or translational symmetries is a fascinating problem, both theoretically and from the
materials point of view
(for a recent review on resonating valence bond phases in frustrated quantum magnets, see
e.g~\cite{misguich_review}).
As discussed by Kanoda at this conference\cite{shimizu_spinliquid},
the compound
$\kappa$-(BEDT-TTF)$_2$Cu$_2$(CN)$_3$ may offer a realization of such a spin-liquid
state (presumably through a combination of strong frustration and
strong charge fluctuations \cite{imada_2003_spinliquid}),
but this behaviour is certainly more the exception than the rule.

\vspace{.6cm}
\noindent
{\bf 2. DYNAMICAL MEAN-FIELD THEORY AND TWO KEY ENERGY SCALES}
\vspace{.4cm}

\begin{figure}
\begin{center}
\begin{tabular}{cc}
\includegraphics[width=\figwidth]{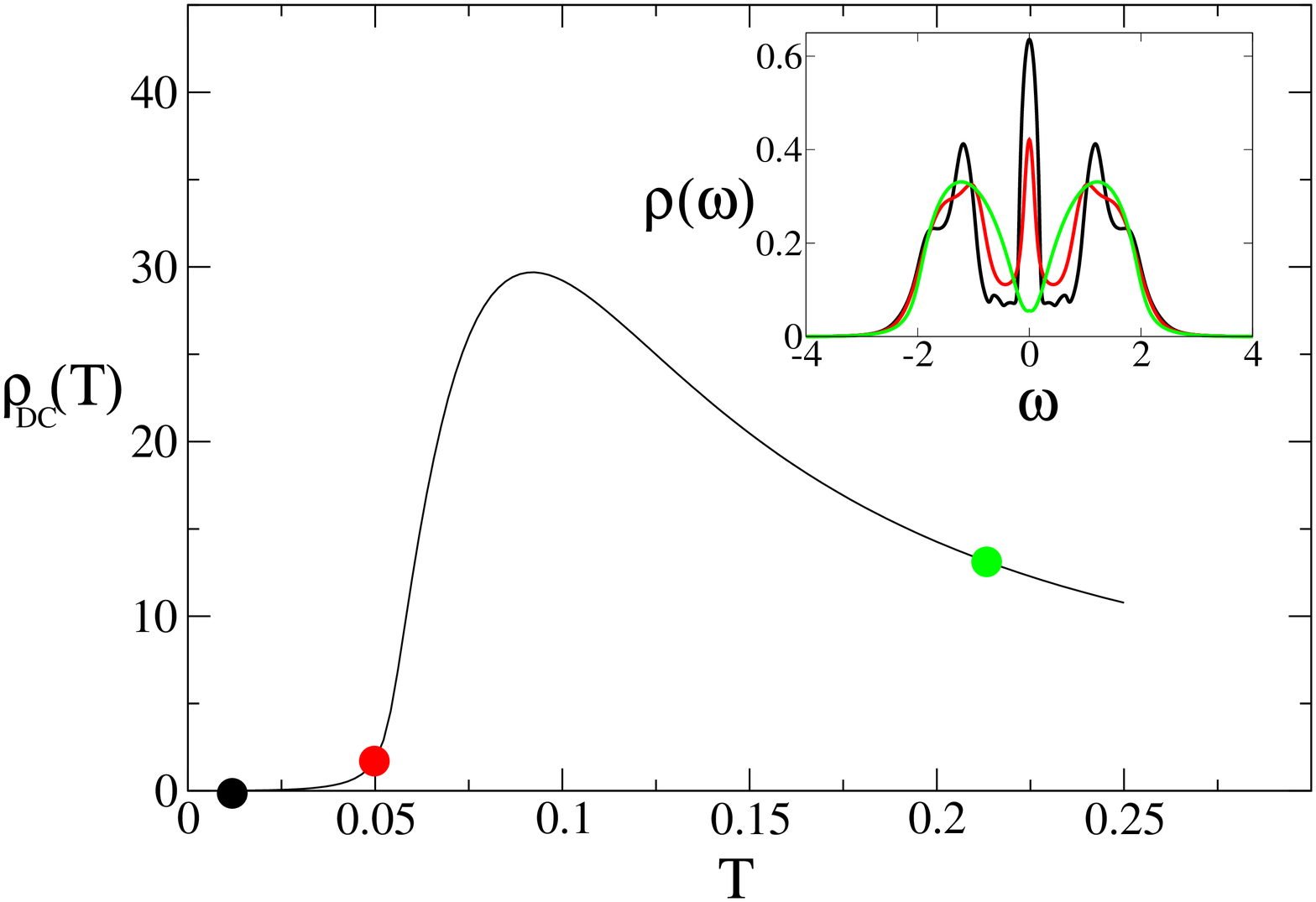} &
\includegraphics[width=\figwidth]{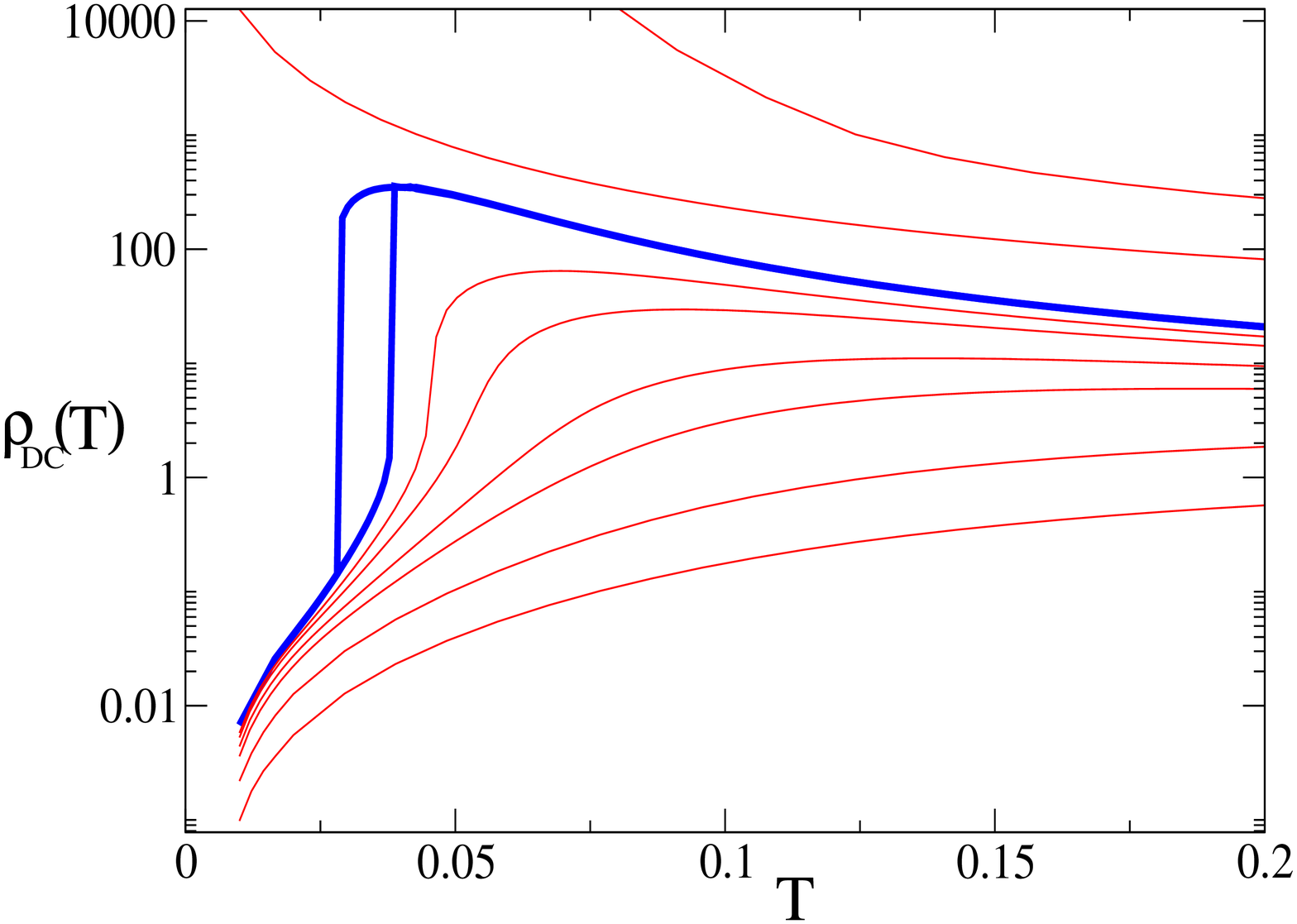}
\end{tabular}
\end{center}
\caption{{\footnotesize
Left: Resistivity in the metallic
phase close to the Mott transition ($U=2.4 D$), as a function of temperature, calculated within
DMFT using the IPT approximation. For three
selected temperatures, corresponding to the three regimes discussed in the
text, the corresponding spectral density is displayed in the inset.
Right: IPT results for the resistivity for values of $U$ in the metallic regime (lower curves),
the coexistence region (bold curve) and the insulating regime (upper two curves).}}
\label{fig:res_and_dos}
\end{figure}

Over the last decade, a detailed theory of the strongly correlated metallic
state, and of the Mott transition itself has emerged,
based on the
{\it dynamical mean-field theory} (DMFT). We refer to \cite{georges_review_dmft} for a review and
an extensive list of original references.
Some key features of this theory are the following~\cite{mott}:

\begin{itemize}
\item In the metallic state, Fermi-liquid theory applies below a low energy scale $\est$,
which can be interpreted as the coherence-scale for quasiparticles. This low-energy coherence scale
is given by $\est\sim Z D$ (with $D$ the half-bandwith, also equal to the Fermi energy
of the non-interacting system at half-filling) where $Z$ is the quasiparticle weight.
In the strongly correlated metal close to the transition, $Z\ll 1$, so that $\est$ is
strongly reduced as compared to the bare Fermi energy.

\item In addition to low-energy quasiparticles (carrying a fraction $Z$ of the spectral weight),
the one-particle spectrum of
the strongly correlated metal contains high-energy excitations
carrying a spectral weight $1-Z$. These are associated to the atomic-like transitions
corresponding to the addition or removal of one electron on an atomic site, which
broaden into Hubbard bands in the solid. As a result, the $\vk$-integrated spectral function
$A(\omega)=\sum_\vk A(\vk,\omega)$ (density of states d.o.s) of the strongly correlated metal
is predicted \cite{georges_kotliar_dmft}
to display a three-peak structure, made of a quasiparticle band close to the Fermi energy surrounded
by lower and upper Hubbard bands (Fig.~\ref{fig:res_and_dos}).
The quasiparticle part of the d.o.s has a reduced width of
order $ZD\sim\est$. The lower and upper Hubbard bands are separated by
an energy scale $\Delta$.

\item At strong enough coupling (see below), the paramagnetic solution of the DMFT equations
is a Mott insulator, with a gap $\Delta$ in the one-particle spectrum. This phase is characterized
by unscreened local moments,
associated with a Curie law for the local susceptibility $\sum_q\chi_q \propto 1/T$, and
an extensive entropy (note however that the uniform susceptiblity $\chi_{q=0}$ is finite, of
order $1/J\sim U/D^2$). As temperature is lowered, these local moments order
into an antiferromagnetic phase~\cite{jarrell}.
The N\'eel temperature is however strongly dependent on frustration~\cite{georges_review_dmft,mott}
(e.g on the ratio $t'/t$ between the next nearest-neighbour and nearest-neighbour hoppings) and can
be made vanishingly small for fully frustrated models.

\end{itemize}

Within DMFT, a separation of energy scales holds close to the Mott transition. The mean-field
solution corresponding to the paramagnetic metal at $T=0$ disappears at a critical coupling
$U_{c2}$. At this point, the quasiparticle weight vanishes ($Z\propto 1-U/U_{c2}$) as in
Brinkman-Rice theory. On the other
hand, a mean-field insulating solution is found for $U>U_{c1}$, with the Mott gap $\Delta$
opening up at this critical coupling (Mott-Hubbard transition).
As a result, $\Delta$ is a finite energy scale for $U=U_{c2}$ and
the quasiparticle peak in the d.o.s is well separated from the Hubbard bands
in the strongly correlated metal.

\begin{figure}
\begin{center}
\includegraphics[width=\figwidth]{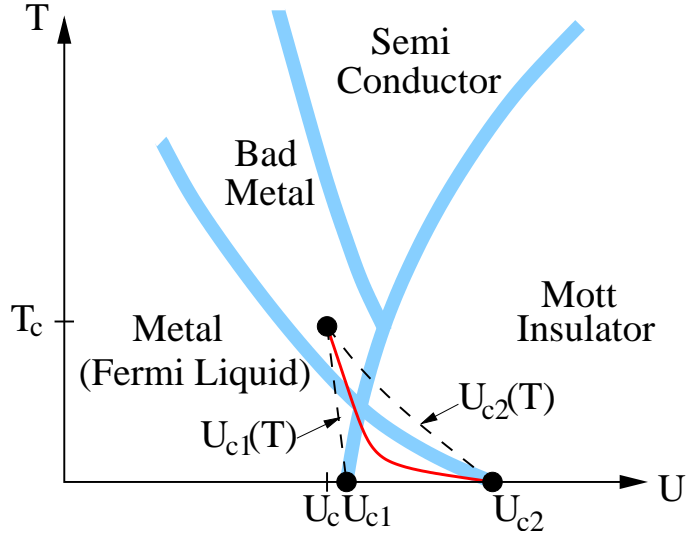}
\end{center}
\caption{{\footnotesize
Paramagnetic phases of the Hubbard model within DMFT,
displaying schematically the spinodal lines of the Mott insulating and
metallic mean-field solutions (dashed),
the first-order transition line (plain) and the critical endpoint. The shaded crossover
lines separating the different transport regimes discussed in Sec.3 are
also shown. The Fermi-liquid to ``bad metal'' crossover line corresponds to the
quasiparticle coherence scale and is a continuation of the spinodal $U_{c2}(T)$
above $T_c$. The crossover into the insulating state corresponds to the continuation
of the $U_{c1}$ spinodal.
Magnetic phases are not displayed and depend on the degree of frustration}}
\label{fig:phasediag}
\end{figure}

These two critical couplings extend at finite temperature into two spinodal lines
$U_{c1}(T)$ and $U_{c2}(T)$, which delimit a region of the $(U/D,T/D)$ parameter space
in which two mean-field solutions (insulating and
metallic) are found (Fig.~\ref{fig:phasediag}). Hence, within DMFT, a first-order
Mott transition occurs at finite temperature even in a purely electronic model.
The corresponding critical temperature $T_c^{el}$ is of order $T_c^{el}\sim \Delta E/\Delta S$, with
$\Delta E$ and $\Delta S\sim \ln (2S+1)$ the energy and entropy differences between the metal
and the insulator. Because the energy difference is small ($\Delta E\sim (U_{c2}-U_{c1})^2/D$), the
critical temperature is much lower than $D$ and $U_c$ (by almost two orders of magnitude).

The existence of well- formed (lower) Hubbard bands in correlated metals was established experimentally more
than ten years ago in the pioneering work \cite{fujimori_pes_oxides}
of Fujimori and coworkers on the photoemission spectra
of $d^1$ transition metal oxides. A clear demonstration of the narrow quasiparticle
peak in $A(\omega)$ predicted by DMFT close to the transition came only recently
however (see \cite{mo_V2O3_prominent_peak} for the metallic phase
of \v2o3 and \cite{matsuura_NiSSe_pes} for NiS$_{2-x}$Se$_x$).
In the case of \v2o3, this was made possible by the use of high-energy photons in
photoemission spectroscopy, allowing to overcome the surface sensitivity of this technique.
This also proved essential in resolving a long-standing controversy on the
spectroscopy of Ca$_{1-x}$Sr$_x$VO$_3$ \cite{maiti_2001,sekiyama_casrvo3}.

The quasiparticle peak in the d.o.s is characterized by an extreme sensitivity to
changes of temperature (inset of Fig.~\ref{fig:res_and_dos}). Its height is strongly reduced as $T$ is increased,
and the peak
disappears altogether as $T$ reaches $\est$, leaving a pseudogap at the Fermi energy. Indeed,
above $\est$, long-lived coherent quasiparticles no longer exist. The corresponding spectral
weight is redistributed over a very large range of energies, of order $U$. These spectral weight
transfers and redistributions are a distinctive feature of strongly correlated systems, and have
been observed e.g in the optical conductivity of both
metallic \v2o3 \cite{rozenberg_optics_prl} and the organics \cite{eldridge_optics_bedt}.
This is reminiscent of Kondo systems \cite{liu_allen_cerium_spectro},
and indeed DMFT establishes a formal and physical connection \cite{georges_kotliar_dmft} between
a metal close to the Mott transition and the Kondo problem. The local moment present at short time-scales
is screened through a self-consistent Kondo process involving the
low-energy part of the (single- component) electronic fluid itself.

\vspace{.6cm}
\noindent
{\bf 3. TRANSPORT REGIMES AND CROSSOVERS}
\vspace{.4cm}

The disappearance of coherent quasiparticles, and associated spectral changes, results
in three distinct transport regimes \cite{rozenberg_optics_prl,majumdar_transport,Mer00,limelette_bedt_prl}
for a correlated metal close to the Mott transition,
within DMFT (Figs.~\ref{fig:phasediag} and \ref{fig:res_and_dos}):

\begin{itemize}

\item In the {\it Fermi-liquid regime} $T\ll \est$, the resistivity obeys a $T^2$
law with an enhanced prefactor:
$\rho = \rho_M\,(T/\est)^2$. In this expression, $\rho_M$
is the Mott-Ioffe-Regel resistivity $\rho_M \propto ha/e^2$ corresponding
to a mean-free path of the order of a single lattice spacing in a Drude picture.

\item For $T\sim\est$, an {\it ``incoherent'' (or ``bad'') metal} regime is entered.
The quasiparticle lifetime shortens dramatically, and the quasiparticle peak is
strongly suppressed (but still present). In this regime, the resistivity is metallic-like (i.e
increases with $T$) but reaches values considerably larger than the Mott ``limit'' $\rho_M$.
A Drude description is no longer applicable in this regime.
%(but such an observation
%does not imply the absence of Fermi-liquid quasiparticles at lower energy).

\item Finally, for $\est\ll T \ll \Delta$, quasiparticles are gone altogether and the d.o.s
displays a pseudogap associated with the scale $\Delta$ and filled with thermal
excitations. This yields a {\it semi-conducting} regime of transport,
with the resistivity decreasing upon heating. Hence, the temperature dependence of
the resistivity displays a maximum (Fig.~\ref{fig:res_and_dos}).
This maximum is observed experimentally in both Cr-doped V$_2$O$_3$ and the organics.

\end{itemize}

\begin{figure}
\begin{center}
\begin{tabular}{cc}
\includegraphics[width=\figwidth]{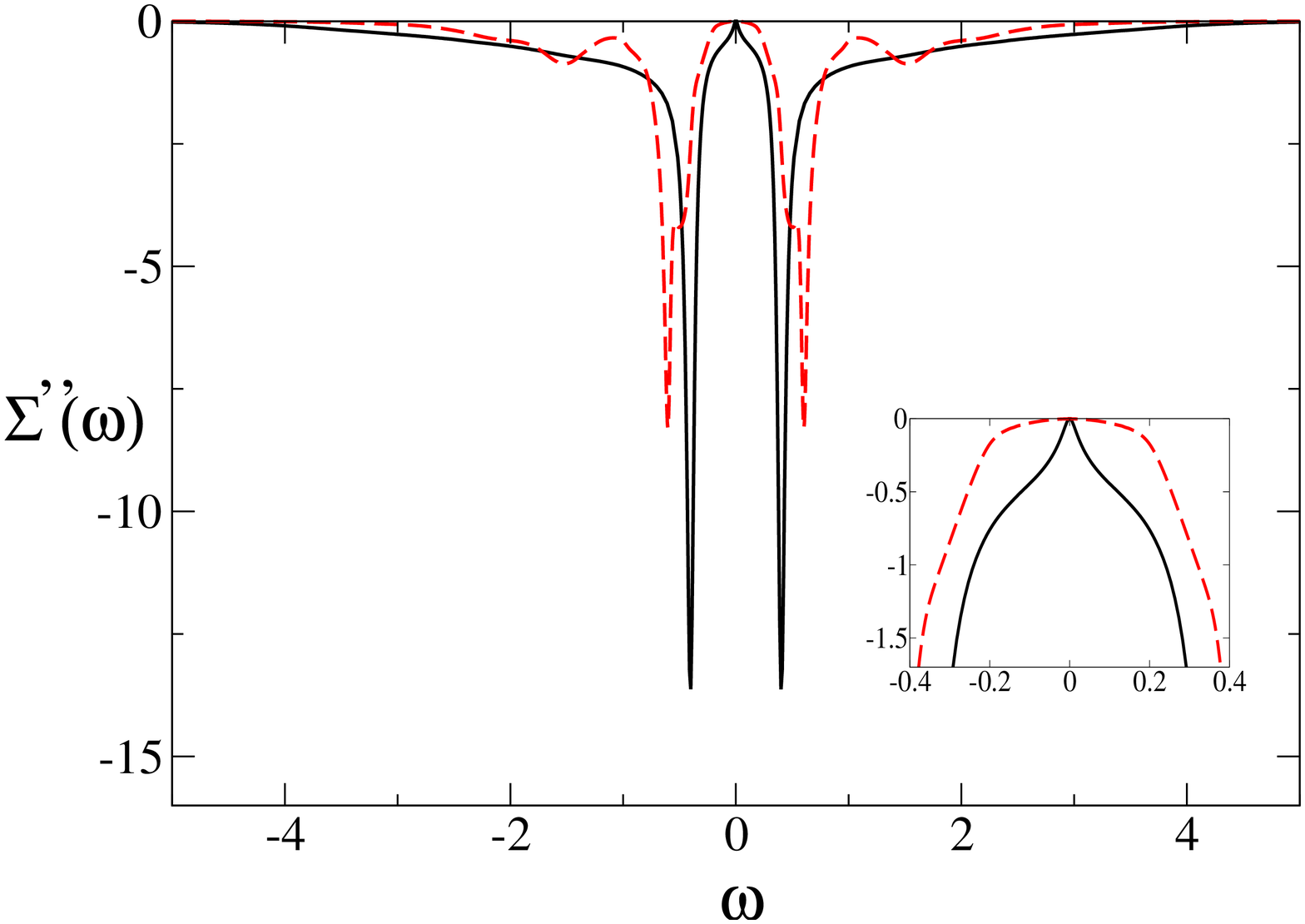} &
\includegraphics[width=\figwidth]{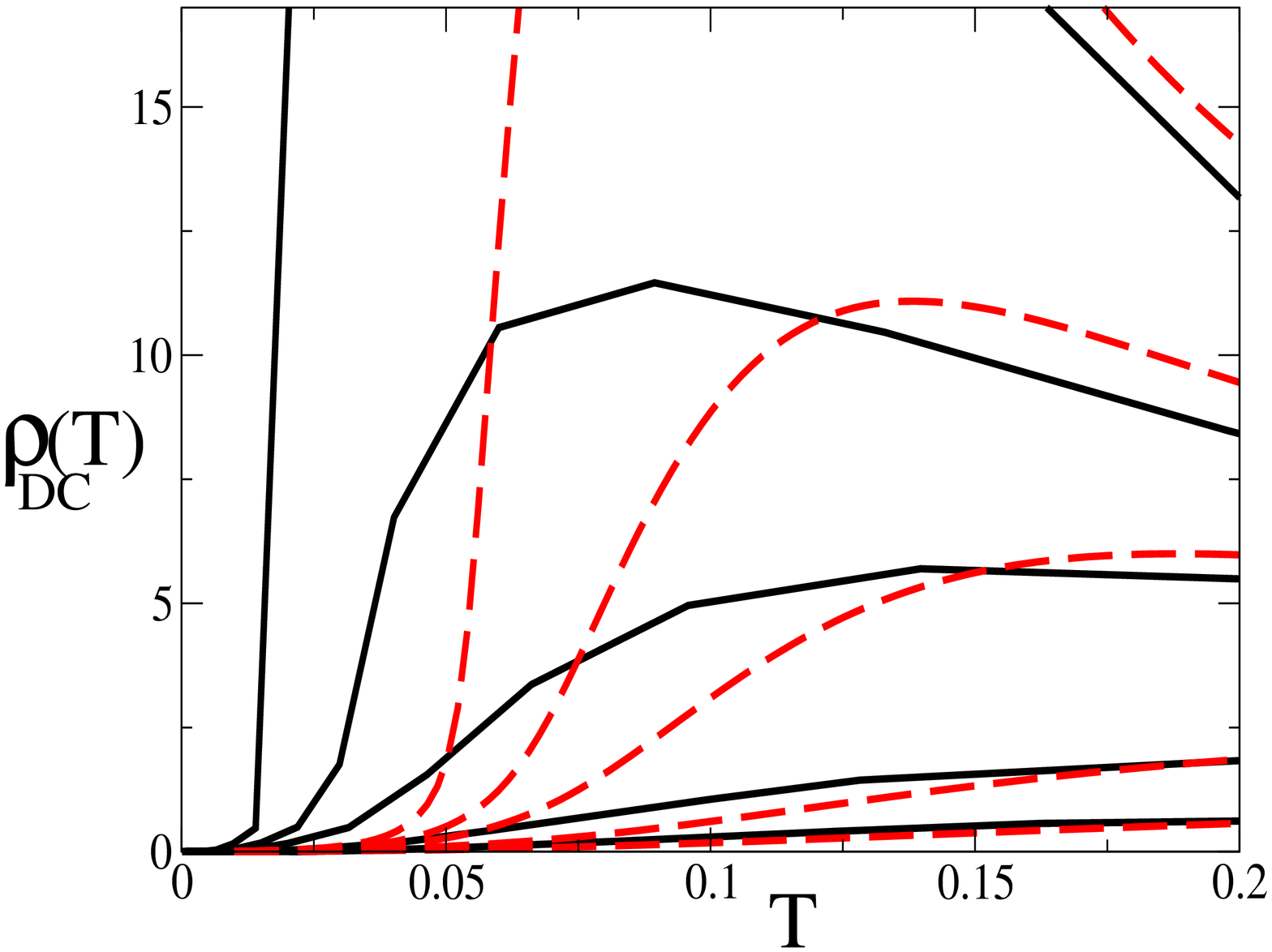}
\end{tabular}
\end{center}
\caption{{\footnotesize
Comparison between the IPT (dashed lines) and NRG methods (plain lines).
Left: The low-frequency behaviour of the inverse lifetime $\mbox{Im}\Sigma$
clearly displays a critically enhanced curvature, which is not reproduced by IPT.
Right: Temperature-dependence of the resitivity in both methods. Note the enhancement of
the $T^2$ term and the lower temperature for the maximum in the NRG results.
}}
\label{fig:compare_ipt_nrg}
\end{figure}

Within DMFT, the conductivity can be simply obtained from a calculation of the
one-particle self-energy since vertex corrections are absent. However, a precise
determination of both the real and imaginary part of the real-frequency self-energy is
required. This is a challenge for most ``impurity solvers''. In practice, early
calculations\cite{majumdar_transport,rozenberg_optics_prl,Mer00}
used the iterated perturbation theory (IPT) approximation\cite{georges_kotliar_dmft}.
The results displayed in
Fig.~\ref{fig:res_and_dos} have been obtained with this technique, and the overall shape of the
resistivity curves are qualitatively reasonable. However, the IPT approximation does a poor
job on the quasiparticle lifetime in the low-temperature regime. Indeed, we expect on general grounds
that, close to the transition, $D\,\mbox{Im}\Sigma$ becomes a scaling function~\cite{moeller_projective}
of $\omega/\est$ and $T/\est$, so that for $T\ll\est$ it behaves as:
$\mbox{Im}\Sigma(\omega=0)\propto D (T/\est)^2\propto T^2/(Z^2D)$ which leads to an
enhancement of the $T^2$ coefficient of the resistivity by $1/Z^2$ as mentioned above.
The IPT approximation does not capture this enhancement and yields the incorrect result
$\mbox{Im}\Sigma_{IPT}(\omega=0)\propto U^2 T^2/D^3$.
For this reason, we have recently used \cite{limelette_bedt_prl}
the numerical renormalization group (NRG) method in order to
perform accurate transport calculations within DMFT. This method is very appropriate in this context,
since it is highly accurate at low energies and
yields real-frequency data\cite{bulla_mott_NRG}.
A comparison of the IPT and NRG results for both the resistivity and lifetime is displayed
in Fig.~\ref{fig:compare_ipt_nrg}. The very different curvatures in the low-T regime in fact
affects the whole T-dependence of the resistivity. In particular, the resistivity
maximum occurs at a temperature much lower than predicted by the IPT approximation.
We found this to be important when comparing these calculations to transport data on
organics\cite{limelette_bedt_prl}. Both these compounds and Cr-doped \v2o3 display the
various crossovers described above.

\vspace{.6cm}
\noindent
{\bf 4. CRITICAL BEHAVIOUR: A LIQUID-GAS TRANSITION
%: THE MOTT CRITICAL ENDPOINT AS A LIQUID-GAS TRANSITION
}
\vspace{.4cm}

Progress has been made recently in identifying the critical behaviour at the Mott
critical endpoint,
both from a theoretical and experimental standpoint.

\begin{figure}
\begin{center}
\includegraphics[width=\figwidth]{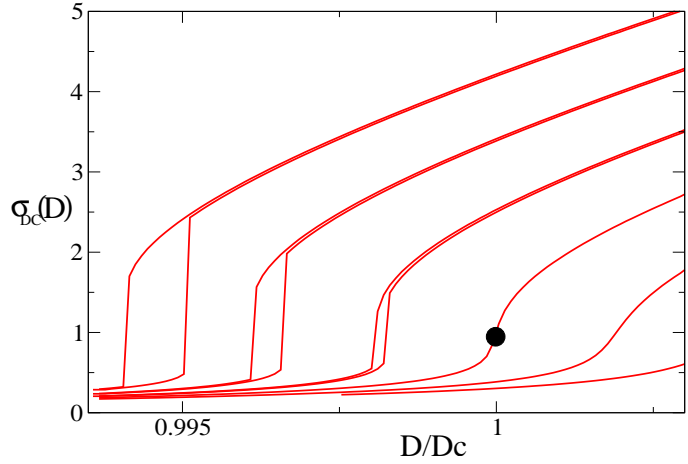}
\end{center}
\caption{{\footnotesize
IPT calculation of the
dc-conductivity as a function of the half-bandwith for the half-filled Hubbard model within DMFT,
for several different temperatures.
Increasing $D$ drives the system more metallic. The curve at $T=T_c$ displays a singularity
(vertical slope: dot)
and hysteretic behaviour is found for $T<T_c$.
}}
\label{fig:sigma_vs_D}
\end{figure}

It has been pointed
early on by Castellani {\it et al.}\cite{Cast79} (see also \cite{Jaya70})
that an analogy exists with the liquid-gas transition in a classical fluid.
Qualitatively, one can focus on the density of double occupancies (or holes)
in each phase:
the insulating phase then corresponds to a low-density ``gas'',
while the metallic phase corresponds to a high-density ``liquid''.
Recently, this analogy has been
given firm theoretical foundations within the framework of a
Landau theory~\cite{kotliar_landau_functional_mott,kotliar_landaumott_prl,rozenberg_finiteT_mott}
derived from DMFT by Kotliar and coworkers.
In this framework, a scalar order parameter $\phi$  is associated with
the low-energy electronic degrees of freedom which build up the quasiparticle
resonance in the strongly correlated metallic phase close to the transition.
This order parameter couples to the singular part of the double occupancy (hence
providing a connection to the qualitative picture above), as well as to other
observables such as the Drude weight or the dc-conductivity.
Because of the scalar nature of the order parameter, the transition falls in
the Ising universality class. In Table 1, the correspondence between the Ising
model quantities, and the physical observables of the liquid-gas transition and of
the Mott metal-insulator transition is summarized.
\begin{center}
{\footnotesize {\bf Table 1:} Corresponding quantities in the liquid-gas
description of the Mott critical endpoint.
The associated Landau free-energy density reads
$r\phi^2+u\phi^4-h\phi$ (a possible $\phi^3$ can be eliminated by an appropriate change
of variables and a shift of $\phi$).} \\
\vspace{.2cm}
\begin{tabular}{|c|c|c|c|} \hline
{\bf Hubbard model} & {\bf Mott MIT} & {\bf Liquid-gas} & {\bf Ising model} \\ \hline
 & & & \\
  $D-D_c$ & $p-p_c$ & $p-p_c$ & Field $h$ \\
  & & & (w/ some admixture of $r$) \\ \hline
  &  &  & Distance to \\
  $T-T_c$ & $T-T_c$ & $T-T_c$ & critical point $r$ \\
  & & & (w/ some admixture of $h$) \\ \hline
  Low-$\omega$ & Low-$\omega$ & $v_g-v_L$ & Order parameter \\
  spectral weight & spectral weight & & (scalar field $\phi$) \\ \hline
\end{tabular}
\end{center}
In Fig.~\ref{fig:sigma_vs_D}, the dc-conductivity obtained from DMFT in the half-filled Hubbard model
(using IPT) is plotted as a function of the half-bandwith $D$, for several different temperatures.
The curves qualitatively resemble those of the Ising model order parameter as a function of
magnetic field (in fact, $D-D_c$ is a linear combination of the field $h$ and of the mass term $r$
in the Ising model field theory). Close to the critical point, scaling implies that
the whole data set can be mapped onto a universal form of the equation of state:
$\langle\phi\rangle= h^{1/\delta}\,f_{\pm}\left(h/|r|^{\gamma\delta/(\delta-1)}\right)$.
In this expression, $\gamma$ and $\delta$ are critical exponents associated with the
order parameter and susceptibility, respectively:
$\langle\phi\rangle\sim h^{1/\delta}$ at $T=T_c$ and
$\chi=d\langle\phi\rangle/dh \sim |T-T_c|^{-\gamma}$.
$f_{\pm}$ are universal scaling functions associated with $T>T_c$ (resp. $T<T_c$).
A quantitative study of the critical behaviour of the double occupancy within DMFT was made
in Ref.~\cite{kotliar_landaumott_prl}, with the expected mean-field values of the
exponents $\gamma=1,\delta=3$.

Precise experimental studies of the critical behaviour at the Mott critical
endpoint have been performed very recently, using a variable pressure technique,
for Cr-doped \v2o3 \cite{limelette_v2o3_science}
and also for the $\kappa$-BEDT organic compounds \cite{kagawa_bedt}. In the case of
\v2o3, a full scaling onto the universal equation of state of the liquid-gas transition
could be obtained \cite{limelette_v2o3_science}.

\vspace{.6cm}
\noindent
{\bf 5. COUPLING TO THE LATTICE, HOT SPOTS AND MORE...}
\vspace{.4cm}

To conclude, we would like to emphasize some open issues and topics of current interest
in connection with the Mott transition and its theoretical description reviewed here.

Lattice degrees of freedom do play a role at the Mott transition in real materials,
e.g the lattice spacing changes discontinuously through the first-order
transition line in (V$_{1-x}$ Cr$_{x}$)$_2$O$_3$. In the metallic
phase, the d-electrons participate in the cohesion of the solid,
hence leading to a smaller lattice spacing than in the insulating phase.
Both the electronic degrees of freedom and the ionic positions must be retained
in order to describe these effects.
In Ref.~\cite{majumdar_compress} (see also~\cite{cyrot_1972_compress},
such a model was treated in the simplest approximation
where all phonon excitations are neglected. The free energy then reads:
$F= K(v-v_0)^2/2 + F_{el}\left[D(v)\right]$. In this expression, $v$ is the unit-cell
volume, $K$ is an elastic constant and the electronic part of the free-energy $F_{el}$ depends on
$v$ through the volume-dependence of the bandwith. In such a model, the critical
endpoint is reached when the electronic response function $\chi=-\partial^2 F_{el}/\partial D^2$
is large enough (but not infinite), and hence the critical temperature $T_c$ of the
compressible model is larger than
$T_c^{el}$ (at which $\chi$ diverges in the Hubbard model). The compressibility
$\kappa = \left(v \partial^2 F/\partial v^2\right)^{-1}$ diverges at $T_c$. This implies
an anomalous lowering of the sound-velocity at the transition\cite{merino_sound},
an effect that has been
experimentally observed in the $\kappa$-BEDT compounds recently \cite{fournier_sound_bedt}.

We emphasize that, within DMFT, a purely electronic model can display
a first-order Mott transition and a finite-T critical endpoint (associated with a diverging
$\chi$), provided that magnetism is frustrated enough so that ordering
does not preempt the transition. Whether this also holds for the finite-dimensional
Hubbard model beyond DMFT is to a large extent an open question
(see \cite{onoda_finiteT_mott} for indications supporting this conclusion
in the 2D case).

There are also important open questions associated with the role of spatial correlations
(inadequately treated by DMFT) in our theoretical understanding of the Mott transition.
In the regime where the quasiparticle coherence scale $\est$ is small as compared
to the (effective strength of the) superexchange $J$, the DMFT picture
is certainly deeply modified. A possibility is that near the transition, the Fermi
surface divides into ``cold'' and ``hot'' regions associated with longer and shorter quasiparticle
lifetimes, respectively (as recently found in a
cluster DMFT study\cite{parcollet_hot}, see also T.Giamarchi {\it et al.} in these proceedings).
Finally, the role of long-wavelength collective modes (in both the charge and spin sectors), and
their feedback on quasiparticle properties is also a key issue which requires to go beyond the
DMFT framework.

\vspace{.6cm}
\noindent {Acknowledgements} \vspace{.4cm}

We are most grateful to the experimental group at Orsay (D. Jerome, P. Limelette, C. Pasquier,
P. Wzietek) as well as to P. Batail, P. Metcalf, C. Meziere and J.M. Honig, for a wonderful
collaboration. We also thank J.Allen, S.R. Hassan, M. Imada, G. Kotliar, H.R. Krishnamurthy,
and M. Rozenberg for numerous discussions.

\noindent
%\bibliographystyle{prsty}
%\bibliography{bibag,patrice}

\end{document}